\documentclass{mem}
\usepackage{natbib}
\usepackage{txfonts}
\usepackage{amssymb}
\usepackage{graphicx}
\usepackage[a4paper]{hyperref}
\idline{1}{1}
\begin{document}
\title{
Markov Chain Monte Carlo methods applied to measuring the fine structure constant from quasar spectroscopy
}

   \subtitle{}

\author{
Julian A. King\inst{1} 
\and Daniel J. Mortlock\inst{2}
\and John K. Webb\inst{1}
\and Michael T. Murphy\inst{3}
 }


\institute{
Department of Astronomy, University of New South Wales,
Sydney, New South Wales, 2052, Australia. \email{jking.phys@gmail.com}
\and
Astrophysics Group, Blackett Laboratory, Price Consort Road, Imperial College London, London SW7
2AZ, U.K.
\and
Centre for Astrophysics and Supercomputing, Swinburne University of Technology,
Victoria, 3122, Australia.
}

\authorrunning{King et al.}

\titlerunning{MCMC methods applied to the fine structure constant in quasar spectra}

\abstract{Recent attempts to constrain cosmological variation in the fine structure constant, $\alpha$, using quasar absorption lines have yielded two statistical samples which initially appear to be inconsistent. One of these samples was subsequently demonstrated to not pass consistency tests; it appears that the optimisation algorithm used to fit the model to the spectra failed. Nevertheless, the results of the other hinge on the robustness of the spectral fitting program VPFIT, which has been tested through simulation but not through direct exploration of the likelihood function. We present the application of Markov Chain Monte Carlo (MCMC) methods to this problem, and demonstrate that VPFIT produces similar values and uncertainties for $\Delta\alpha/\alpha$, the fractional change in the fine structure constant, as our MCMC algorithm, and thus that VPFIT is reliable.
}
\maketitle{}

\section{Introduction}

Recent years have seen sustained interested in attempting to determine whether any of the fundamental constants of nature vary. Efforts have been focused in particular on the fine structure constant, $\alpha$, and the proton-to-electron mass ratio, $\mu$, although other dimensionless constants have also been considered. Quasars provide a method of probing the value of $\alpha$ in the early universe by illuminating gas clouds along the line of sight to Earth. In particular, certain absorption lines in these gas clouds are sensitive to changes in one or more of the fundamental constants of nature. For an atom/ion, the relativistic corrections to the energy levels of an electron are proportional to $\alpha^2$, although the magnitude of the change depends on the transition under consideration. By comparing the wavelengths of absorption lines with differing sensitivities to a change in $\alpha$, we are able to place constraints on a change in $\alpha$ between the early universe and today. That is, we are able to measure $\Delta\alpha/\alpha = (\alpha_z-\alpha_0)/\alpha_0$, where $\alpha_z$ is the value of $\alpha$ at redshift $z$ and $\alpha_0$ is the laboratory value.

The tentative detection of a variation in $\alpha$ reported by \citet{Webb99} has further increased interest in this field. Subsequent efforts refined this work by increasing the number of absorption systems considered. This yielded a constraint of $\Delta\alpha/\alpha = (-0.57 \pm 0.11)\times 10^{-5}$ \citep{Murphy04} from 143 absorption systems. However all of these observations are from the Keck/HIRES (High Resolution Echelle Spectrometer) instrument, and so it remains important to confirm such unexpected results with independent equipment. 

\citet{Chand04} reported $\Delta\alpha/\alpha = (-0.06 \pm 0.06) \times 10^{-5}$ from 23 measurements using the Ultraviolet and Visual Echelle Spectrograph (UVES), on the Very Large Telescope (VLT). These two measurements of $\Delta\alpha/\alpha$ are clearly discrepant. A later analysis \citep{Murphy08a} found that the analysis of \citeauthor{Chand04} cannot be correct as it states a statistical precision in excess of the maximum theoretical precision allowed by the data. Similarly, \citeauthor{Murphy08a} analysed the $\chi^2$ vs $\Delta\alpha/\alpha$ curves produced by the \citeauthor{Chand04} optimisation algorithm, and concluded that the shape of the curves demonstrate a failure of the algorithm to find the maximum likelihood value of $\Delta\alpha/\alpha$ implied by the model fits and data, and thus that the estimate of $\Delta\alpha/\alpha$ given by \citeauthor{Chand04} is unreliable. 

Although it would appear that the \citeauthor{Murphy04} results are robust, it is worth directly investigating the optimisation algorithm used in order to confirm that it is reliable. Furthermore, each measurement of $\Delta\alpha/\alpha$ will be subject to systematic errors, but some systematic errors should have an expectation value of zero, and thus averaging over many absorption systems will in principle eliminate such errors. It has become commonplace to quote values of $\Delta\alpha/\alpha$ for individual systems; in these cases, one has no method of determining the size of certain systematic errors through comparison with other systems, and so one should be particularly careful that the statistical errors are stated correctly. The MCMC method we describe herein allows one to confirm the validity of the statistical errors produced by a different method of analysis.  

\section{Motivation for MCMC}

Ideally one would like an independent method of demonstrating whether or not a purported value of $\Delta\alpha/\alpha$ and the associated statistical uncertainty given by an optimisation algorithm are reliable. 

Optimisation algorithms seek to minimise $\chi^2 = \sum_i [I(\mathbf{x})_i - d_i]^2 / \sigma_i^2$, where $d_i$ are the spectroscopic data points, $I(\mathbf{x}_i)$ is the prediction of the model at each data point, and $\sigma_i$ is the standard error associated with each spectroscopic data point. However, $\chi^2 = -2 \ln(L(\mathbf{x}))$ up to an additive constant which can be neglected, as we only ever consider differences of $\chi^2$ for finding model parameters. We thus use the definition of the likelihood function $L(\mathbf{x}) \equiv \exp[-\chi^2(\mathbf{x})/2]$.

One option is to use an alternate optimisation algorithm. Although optimisation algorithms are in principle simple, numerical issues can cause them to fail inconsistently; this may be the case for the algorithm utilised by \citet{Chand04}. In particular, the optimisation algorithms employed by \citet{Chand04}, \citet{Murphy08a} and \citet{Murphy04} are of the Newton type, which require all first and second partial derivatives of $\chi^2$ with respect to to the parameters to be known. The Voigt function used to model the absorption lines is not analytic, and nor are its derivatives. As such, partial derivatives must be approximated by finite difference methods. Inappropriate choices of the step size for the finite differencing scheme can either produce poor approximations to the derivatives (step size too large) or be rendered useless by roundoff error (step size too small), leading to poor performance of the optimisation algorithm. There are number of other numerical issues which may cause failure of the optimisation algorithm, but we do not consider these here.

On account of these numerical issues, one would desire to explore the parameter space itself to directly determine the confidence limits on $\Delta\alpha/\alpha$. 

\section{Description of the MCMC method}

Traditional Monte Carlo methods suffer from the ``curse of dimensionality''. That is, their performance degrades exponentially with increasing dimensionality. The Markov Chain Monte Carlo (MCMC) method degrades only polynomially with increased dimensionality, at the expense of introducing correlation between samples. Additionally, MCMC methods must be tuned to the probability distribution under consideration so as to explore the parameter space efficiently. 

We implement a variant of the Metropolis algorithm \citep{Metropolis1953} to explore our parameter space. The Metropolis algorithm proposes a new position in the parameter space, $\mathbf{x}'$, based on the current position, $\mathbf{x}$, according to some proposal function, $T(\mathbf{x},\mathbf{x}')$. The only requirement imposed is that $T(\mathbf{x},\mathbf{x}') = T(\mathbf{x}',\mathbf{x})$ (i.e the proposal distribution is symmetric). 

Although in principle there are large numbers of possible proposal funcitons, $T$, in practice the most common choice is a multidimensional Gaussian centred on the current point, such that $\mathbf{x}' = \mathbf{x} + gN(0,\mathbf{\Sigma})$ where $\mathbf{\Sigma}$ is the covariance matrix obtained from the optimisation algorithm at the purported best fit solution, and $g$ is a scalar tuning factor. The choice of $T$ influences only the efficiency of the algorithm, not the formal correctness of the solution. The initial $\mathbf{\Sigma}$ may or may not be a good approximation to the true convariance matrix. The use of $\mathbf{\Sigma}$ ensures that the distribution of proposed parameters is approximately the same as the underlying distribution; the closer $\mathbf{\Sigma}$ is to the true covariance matrix, the faster the MCMC algorithm will be. 

The tuning factor $g$ effectively controls the size of steps taken. If $g$ is too large, most trial steps will land in regions of low likelihood, and therefore most steps will be rejected (the chain will not move). On the other hand, if $g$ is too small, the acceptance rate will be $\approx 100\%$, but the parameter space will be explored too slowly. If both the target and proposal distributions are Gaussian then the ideal acceptance rate is $\approx 44\%$ \citep{GRG95}. 

The algorithm generates a sequence of points, $\{\mathbf{x}^t\}$, according to a two step prescription. First, from the current point, $\mathbf{x}$, propose a new point, $\mathbf{x}'$, via $T(\mathbf{x},\mathbf{x}')$. Then calculate the ratio $q=L(\mathbf{x}')/L(\mathbf{x})$. Secondly, with probability $\min(q,1)$ move to the new point i.e. set $\mathbf{x}^{t+1}=\mathbf{x}'$. Otherwise, retain the current point i.e. $\mathbf{x}^{t+1}=\mathbf{x}^t$. In this fashion, proposed moves to a point which is more likely than the existing point are always accepted, whereas proposed moves to a point which is less likely than the existing point are sometimes accepted, depending on the ratio of likelihoods. For a sufficiently large numbers of iterations, and with proper tuning of the algorithm, the distribution of $\{\mathbf{x}^t\}$ will sample from the underlying probability distribution, up to a normalisation constant. In particular, $\{\mathbf{x}^t_{\Delta\alpha/\alpha}\}$ will sample from the probability distribution of $\Delta\alpha/\alpha$, from which we can obtain a best estimate and confidence limits.

To minimise running time, for each model fit we run our MCMC algorithm several times (usually five to ten, depending on the complexity of the situation, with several hundred thousand iterations per stage) and re-estimate $\mathbf{\Sigma}$ at each stage from the chain. Prior to starting each MCMC run, we execute small runs (typically 250 iterations) in which we tune $g$ to be such that the acceptance rate is between $30\%$ and $50\%$ (the outputs from these small runs do not count towards each stage). Thus, even if the initial covariance matrix does not allow a good exploration of the parameter space, by re-estimating $\mathbf{\Sigma}$ and retuning $g$ several times we can drastically increase the chance that the final MCMC run will produce a good approximation of the underlying probability distribution. We determine whether the final MCMC run is useful by examining the chain for autocorrelation. If the autocorrelation length in the chain is much smaller than the chain length, then we deem the final run acceptable.

We do not require the usual ``burn-in'' period, where one discards a certain number of samples from the start of the chain, because we believe our parameters already start at the likelihood maximum. We can determine whether this assumption is robust by examining the chain -- parameters should stay near their starting values, on average, if our initial parameter estimates were good.

We implement the Multiple Try Metropolis algorithm \citep{Liu2000}, which expands the Metropolis algorithm to allow multiple attempts at each step. If the initial proposal distribution is poorly tuned, this variant of the Metropolis algorithm tends to be much more robust for larger number of dimensions. 

Similarly, we do not use a Gaussian proposal distribution, but start with a radial distribution which has $P(r) \propto (2/3)r^2\exp(-r^2/2) + (1/3)\exp(-r)$. This mixture of an exponential distribution and the radial component of a 2D Gaussian allows the algorithm to occasionally take large steps, whilst otherwise taking steps clustered about some value; this speeds exploration of the parameter space where $\mathbf{\Sigma}$ is initially poorly tuned. For our proposal distribution, $T$, we generate our parameters from a spherically symmetric distribution with radial probability density $P(r)$ and then left multiply by $\mathbf{L}$ (where $\mathbf{LL^T}=\mathbf{\Sigma}$) so that the proposal distribution has the correct covariance structure.

MCMC can be used to directly estimate posterior probabilities in the Bayesian framework with the appropriate choice of prior distribution. The likelihood ratio then becomes $L(\mathbf{x}) \rightarrow L(\mathbf{x})\pi(\mathbf{x})$, where $\pi(\mathbf{x})$ is the Bayesian prior for a particular set of parameters. We utilise improper flat priors for the column densities and redshifts of each component. Similarly, we utilise a flat prior on $\Delta\alpha/\alpha$. 

We utilise a flat prior for the logarithm of the Doppler parameters, rather than the Doppler parameters, to suppress movements to small $b$. Otherwise, the algorithm tends to propose many jumps to $b<0$ for narrow lines, which must be rejected -- this substantially reduces the efficiency of the algorithm. This is somewhat reasonable also on physical grounds, as we do not expect large numbers of gas clouds described by arbitrarily small temperatures, and certainly our fitting procedure would reject a profile decomposition of this nature. Using a flat prior for the Doppler parameters results in unacceptably large running times, and so we use the logarithmic prior as an easy and practical option.

We have modified the spectral profile fitting program VPFIT\footnote{See http://www.ast.cam.ac.uk/$\sim$rfc/vpfit.html} to incorporate our MCMC algorithm. The outputs of the optimisation algorithm are fed directly into the initialisation for the MCMC code. Our MCMC algorithm uses the same code as VPFIT to generate the Voigt profiles and calculate $\chi^2$, and thus our algorithm does not eliminate the possibilty of a code bug entirely. However, with this caveat, our algorithm can determine whether the optimisation code used by VPFIT does or does not converge to the desired solution and produce appropriate uncertainty estimates.

\section{Results}

\begin{table*}
\caption{\label{tabresults}Comparison of purported values of $\Delta\alpha/\alpha$ calculated by VPFIT, and the results of the MCMC algorithm. Quoted uncertainties are $1\sigma$.}
  
\begin{center}
\begin{tabular}{llll}\hline
Object & Redshift & $\Delta\alpha/\alpha$ -- VPFIT & $\Delta\alpha/\alpha$ -- MCMC \\\hline
LBQS\, 2206$-$1958 & 1.018 & $(-0.51 \pm 1.07) \times 10^{-5}$ & $(-0.51 \pm 0.88) \times 10^{-5}$\\
LBQS\, 0013$-$0029 & 2.029 & $(-0.86 \pm 0.94) \times 10^{-5}$ & $(-0.83 \pm 0.77) \times 10^{-5}$\\
Q\,0551$-$366 & 1.748 & $(-0.80 \pm 1.08) \times 10^{-5}$ & $(-0.89 \pm 0.84) \times 10^{-5}$\\\hline
  \end{tabular}
  \end{center}
 
\end{table*} 

We have applied our MCMC algorithm as described above to the three quasar absorption systems described below. The numerical values produced by the optimisation algorithm (``VPFIT'') and the MCMC code (``MCMC'') are given in table \ref{tabresults}. In all cases we find good agreement between the VPFIT result and that produced by our MCMC code, although the statistical uncertainties produced by our MCMC code are mildly smaller than that produced by VPFIT, indicating that VPFIT may be conservative.

Our fits all pass appropriate robustness tests (in particular, $\chi^2_\nu \approx 1$ where $\nu$ is the number of degrees of freedom for the fit). All of our final chains mix well, although with the second and third objects considered here the initial chains do not -- re-estimation of the covariance matrix multiple times is necessary to achieve a well mixed chain. All redshifts here refer to the redshift of the absorption system.

\subsection{LBQS\,2206$-$1958 $z=1.018$}

This absorption system appears to be well fitted by a single Voigt profile. We use the Mg{\sc \,ii} $\lambda\lambda2796,2803\AA$ transitions, which are relatively insensitive to $\alpha$ variation, and the Fe{\sc \,ii} $\lambda \lambda \lambda \lambda 2382,2600,2344,2587\AA$ transitions, which are strongly sensitive.  

The parameters are approximately jointly normally distributed. We expect this for single component fits -- the Voigt profile decomposition is effectively unique with one component.

\subsection{LBQS\,0013$-$0029 $z=2.029$}

This system appears with two obvious absorption features. We find that the bluer feature is better fitted by two components than it is by one on the basis of a statistically significant reduction in $\chi^2$ when using two components. That is, we fit three components to this absorption profile. We use a wide variety of transitions, namely: Si{\sc \,ii} $\lambda1526\AA$, Al{\sc \,iii} $\lambda \lambda 1854, 1862\AA$, Fe{\sc \,ii} $\lambda \lambda \lambda \lambda \lambda 2382,2600,2344,2587,1608\AA$ and Mg{\sc \,i} $\lambda 2852\AA$.  

The chain values of $\Delta\alpha/\alpha$ are approximately Gaussian-distributed.

\subsection{Q\,0551$-$366 $z=1.748$}

\begin{figure}
\begin{center}
\resizebox{\hsize}{!}{\includegraphics[hiresbb,viewport=1 1 92 151,angle=-90,width=61mm]{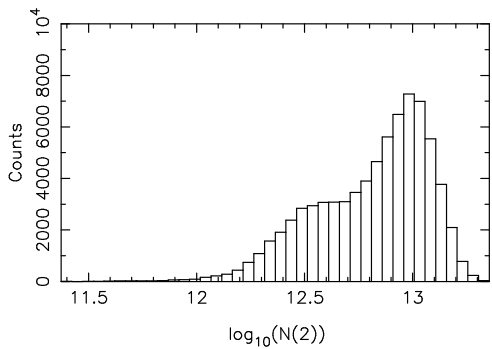}}
\end{center}
\caption{\label{figQ0551366a}Histogram of the chain for $\log_{10}(N(2)/\rm{cm}^2)$, where $N(2)$ is the column density of the central component to the fit for the $z=1.748$ absorption system toward Q\,0551$-$366.}
\end{figure}

\begin{figure}
\begin{center}
\resizebox{\hsize}{!}{\includegraphics[hiresbb,viewport=1 1 90 148,angle=-90,width=61mm]{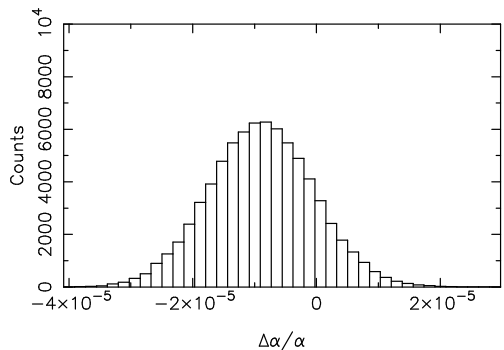}}
\end{center}
\caption{\label{figQ0551366b}Histogram of the chain for $\Delta\alpha/\alpha$ for the $z=1.748$ absorption system toward Q\,0551$-$366.}
\end{figure}

This absorption feature appears as one weak feature next to one relatively strong feature, with some overlap. We find the bluer feature to be well modelled by one component, however the higher wavelength feature appears to require two closely spaced components to achieve a statistically acceptable fit. Hence, we use three components in total to model the observed profile. The transitions we use are: Si{\sc \,ii} $\lambda1526\AA$, Mg{\sc \,i} $\lambda 2852\AA$ and Fe{\sc \,ii} $\lambda \lambda \lambda \lambda \lambda \lambda$ $2382,2600,2344,2587,1608,2374\AA$.

We note that the parameters corresponding to the two highest wavelength components are manifestly not normally distributed (see Fig.\,\ref{figQ0551366a} for an example). We confirm, by inspection of the chain, that this effect is not due to permutations of corresponding parameters, which would leave $\chi^2$ unchanged. 

Despite the gross departures from Gaussianity for the column density and Doppler parameters corresponding to the two reddest components, the histogram of $\Delta\alpha/\alpha$ remains approximately Gaussian (see Fig.\,\ref{figQ0551366b}). The Voigt profile decomposition is not unique, and so we can only try to find the model which best describes the observations statistically. However, for a given model we would naively expect that there is a unique value of $\Delta\alpha/\alpha$ which minimises $\chi^2$, and additionally that $\Delta\alpha/\alpha$ should be approximately Gaussian. We find both of these statements to be true here. 

It is reassuring that we find concordance between the VPFIT and MCMC results for $\Delta\alpha/\alpha$ given the significant non-Gaussianity in some parameters. For non-Gaussian parameters, the parameter estimates produced by VPFIT will be the correct maximum likelihood estimates, however the confidence intervals will be biased. For our present purposes, we are only interested in the confidence limits on $\Delta\alpha/\alpha$, and here we find an acceptable level of agreement.

\subsection{Combination of results}

If we assume that a single value of $\alpha$ underlies these, we can combine the three VPFIT results above using a weighted mean to estimate $\Delta\alpha/\alpha = (-0.74 \pm 0.59) \times 10^{-5}$, which is statistically consistent with no change in $\alpha$. We use the VPFIT results as a more conservative estimate.

\section{Discussion \& conclusion}

Given suitable knowledge of the observed distribution of column densities and Doppler parameters, we could implement these distributions as priors to our model. However, our statistical constraints are generally sufficiently good that this should not significantly alter our parameter estimates. In any event, we are primarily interested in $\Delta\alpha/\alpha$, for which a flat prior is reasonable on ignorance grounds. 

We would like to apply our algorithm to verify the results of \citet{King08} in relation to $\Delta\mu/\mu$; however, those fits involve greater than a thousand parameters each. In this context, our algorithm is hopelessly inadequate. Our running times for the objects described herein varied from a few hours to a few days, and have only a few tens of parameters. Exploration of more complicated cases must wait for advances in computing power.

Our results demonstrate that VPFIT produces reliable parameters estimates and uncertainties for relatively simple situations. Experience with VPFIT suggests that there does not appear to be any indication of failure with moderately complicated circumstances, and so we would argue that the optimisation algorithm used by VPFIT is robust. The implication of is this it that the results of \citet{Murphy04} are unlikely to be explained by some failure of the optimisation algorithm used by VPFIT. Thus the detection of a change in $\alpha$ must either be real, or due to some other unknown issue.  

\begin{acknowledgements}
Presentation of this work at the 2009 IAU XXVII General Assembly JD9 conference was supported through financial assistance from the UNSW PRSS scheme.\end{acknowledgements}

\bibliographystyle{aa}
\bibliography{SAIT_KING.bib}
\end{document}